\documentclass[11pt]{article}
\DeclareMathAlphabet{\scr}{U}{rsfs}{m}{n}
\usepackage{latexsym}
\usepackage{epsfig}
\usepackage[mathscr]{eucal}
\usepackage{amsfonts}
\usepackage{amscd}
\usepackage{amsmath}
\usepackage{array}	
\usepackage{amssymb}
\usepackage{colordvi}
\usepackage{enumerate}
\usepackage{graphicx}
\usepackage{booktabs}
\usepackage[footnotesize]{caption}
\usepackage{fancyhdr}
\usepackage{pdfpages}
\usepackage{slashed}
\usepackage{tabularx}
\usepackage{longtable}
\usepackage{array}

\newcommand{\eq}[1]{\begin{equation} #1 \end{equation}}
\newcommand{\eqa}[1]{\begin{eqnarray} #1 \end{eqnarray}}

\newcommand{\newc}{\newcommand}
\newc{\DM}{dark matter\;}
\newc{\KK}{Kaluza-Klein\;}
\newc{\ff}{fragmentation function\;}
\newc{\be}{\begin{equation}}
\newc{\ee}{\end{equation}}
\newc{\bi}{\begin{itemize}}
\newc{\ei}{\end{itemize}}
\newc{\benu}{\begin{enumerate}}
\newc{\eenu}{\end{enumerate}}
\newc{\bc}{\begin{center}}
\newc{\ec}{\end{center}}
\newc{\bfig}{\begin{figure}}
\newc{\efig}{\end{figure}}
\newc{\neutone}{\tilde{\chi}^0_1}

\begin{document}

\title{\hfill ~\\[-30mm]
\phantom{h} \hfill\mbox{\small SLAC-PUB-16088} \\[-1.1cm]
\phantom{h} \hfill\mbox{\small SFB/CPP--14--74} \\[-1.1cm]
\phantom{h} \hfill\mbox{\small TTK--14--23}
\\[1cm]
\vspace{13mm}   \textbf{Electroweak fragmentation functions for dark matter annihilation}}
\date{}
\author{
Leila Ali Cavasonza$^{1}$\footnote{E-mail: \texttt{cavasonza@physik.rwth-aachen.de}}\;,
Michael Kr\"amer$^{1,2\,}$\footnote{E-mail:
  \texttt{mkraemer@physik.rwth-aachen.de}} \, and
Mathieu Pellen$^{1\,}$\footnote{E-mail: \texttt{pellen@physik.rwth-aachen.de}}\\[9mm]
{\small\it
$^1$Institute for Theoretical Particle Physics and Cosmology,}\\ {\small \it RWTH Aachen University, D-52056 Aachen, Germany}\\[1mm]
{\small\it
$^2$ SLAC National Accelerator Laboratory, Stanford University, Stanford, CA 94025, USA}
}

\maketitle

\begin{abstract}

Electroweak corrections can play a crucial role in dark matter annihilation. The emission of gauge bosons, in particular, leads to a secondary flux consisting
of all Standard Model particles, and may be described by electroweak fragmentation functions.
To assess the quality of the fragmentation function approximation to electroweak radiation in dark matter annihilation, we have calculated the flux of secondary particles from gauge-boson emission in
models with Majorana fermion and vector dark matter, respectively. For both models, we have compared cross sections and energy spectra of positrons
and antiprotons after propagation through the galactic halo in the fragmentation function approximation and in the full calculation. Fragmentation functions fail to describe
the particle fluxes in the case of Majorana fermion annihilation into light fermions: the helicity suppression of the lowest-order cross section in such models
cannot be lifted by the leading logarithmic contributions included in the fragmentation function approach. However, for other classes of models like
vector dark matter, where the lowest-order cross section is not suppressed, electroweak fragmentation functions provide a simple, model-independent and accurate description of secondary particle fluxes.
\end{abstract}
\thispagestyle{empty}
\vfill
\newpage
\setcounter{page}{1}

\tableofcontents

\newpage

\section{Introduction}
\label{ch:introduction}

The existence of \DM\cite{Bertone:2004pz} provides strong evidence for physics beyond the Standard Model (SM).
A leading candidate for dark matter are weakly interacting massive particles, which may be produced at colliders or
detected through direct and indirect detection experiments. In indirect detection, in particular, one searches for \DM annihilation products,
including antimatter particles like positrons and antiprotons, which propagate through the galactic halo and which can be detected in astrophysical experiments at the earth.

Electroweak (EW) corrections may be important for dark matter annihilation for two reasons~\cite{Barbot:2002ep, Berezinsky:2002hq, Barbot:2002gt, Kachelriess:2007aj, Bell:2008ey, Dent:2008qy, Barger:2009xe, Fortin:2009rq, Kachelriess:2009zy, Ciafaloni:2010qr, Ciafaloni:2010ti, Bell:2010ei, Bell:2011eu, Bell:2011if,  Ciafaloni:2011sa, Garny:2011cj, Garny:2011ii, Ciafaloni:2012gs, Bell:2012dk,  Ciafaloni:2013hya, Bringmann:2013oja, Ibarra:2014vya, Baratella:2013fya}.
First, the radiation of a vector boson can lift the helicity suppression of cross sections for Majorana fermion dark matter annihilating into light 
fermions~\cite{Bell:2010ei, Bell:2011eu, Bell:2011if, Ciafaloni:2011sa, Goldberg:1983nd, Go832,Bergstrom:1989jr, Bringmann:2007nk, Bergstrom:2008gr}.
Moreover, the emission and decay of electroweak gauge bosons from the primary annihilation products alter the spectrum and composition of the
secondary flux. In particular, the gauge boson decay will lead to a secondary flux which includes all stable SM particles ($e^+, e^-, \nu, \bar{\nu}, \gamma, p, \bar{p}$),
irrespective of the model-specific composition of the primary annihilation products.

Many models provide \DM candidates with masses in the TeV-range, see \textit{e.g.}\ Ref.~\cite{Bertone:2004pz, Jungman:1995df, Cahill-Rowley:2013gca, Cahill-Rowley:2014boa}. For such heavy dark matter, soft and collinear electroweak
gauge boson emission from the relativistic final-state particles is enhanced by Sudakov logarithms $\ln^2 (M^2_{\rm DM}/M^2_{\rm EW})$~\cite{Ciafaloni:1998xg,
Kuhn:1999nn, Fadin:1999bq}, where $M_\text{DM}$ and $M_\text{EW}$ are the mass
of the \DM candidate and of the electroweak gauge boson, respectively.  The leading logarithmic contributions to the annihilation cross section can be described in a model-independent
way by electroweak fragmentation functions~\cite{Ciafaloni:2010ti, Ciafaloni:2001mu, Ciafaloni:2005fm}, where improved splitting functions are introduced in order to take into account the masses of the emitted gauge bosons.

The purpose of the present article is to examine the quality of the \ff approximation. To this end we have compared the
predictions obtained for the secondary flux after propagation using the \ff approximation against those obtained from an
exact calculation. To perform the comparison we have chosen two specific \DM models, a
simplified version of the minimal supersymmetric model (MSSM)~\cite{Jungman:1995df, Ellis:1983ew} with neutralino dark matter,
and a model with universal extra dimensions (UED) where the first Kaluza-Klein excitation of the photon provides a dark matter candidate~\cite{Cheng:2002ej,Servant:2002aq}.
Both models are generic for the annihilation of Majorana fermion and vector dark matter, respectively. To assess the quality of the \ff approach we focus on the
particular case where the \DM particles annihilate at lowest order into electron-positron pairs only.
We show that the \ff approach reproduces well the exact result in the case of UED with vector dark matter, while the approximation
does not work for the MSSM with Majorana fermion annihilation into electron-positron pairs. This is due to the fact that the annihilation of
Majorana fermions into a light lepton pair is helicity suppressed and that the
emission of soft and collinear gauge bosons from the final-state particles, included in the \ff approximation, is not sufficient to lift this
helicity suppression \cite{Ciafaloni:2011sa}. Hence, the \ff approximation provides a
simple and model-independent way to obtain realistic predictions for \DM indirect detection for those models where the
annihilation cross section is not suppressed at the lowest order.

This article is organised as follows: in Section~\ref{ch:frag_function}
we briefly review the \ff approximation to electroweak radiation.
In Section~\ref{ch:models_setup} we present the calculation of EW gauge boson emission in two specific models with Majorana fermion and vector dark matter, respectively, and compare the \ff
approximation against the exact result for the primary flux.
In Section~\ref{ch:pythia} we perform the comparison for the secondary flux after propagation through the galactic halo. We conclude in Section~\ref{ch:conclusions}.

\section{Electroweak corrections}
\label{ch:frag_function}

Electroweak corrections in the context of \DM annihilation have been widely discussed in the literature~\cite{Barbot:2002ep, Berezinsky:2002hq, Barbot:2002gt, Kachelriess:2007aj, Bell:2008ey, Dent:2008qy, Barger:2009xe, Fortin:2009rq, Kachelriess:2009zy, Ciafaloni:2010qr, Ciafaloni:2010ti, Bell:2010ei, Bell:2011eu, Bell:2011if,  Ciafaloni:2011sa, Garny:2011cj, Garny:2011ii, Ciafaloni:2012gs, Bell:2012dk,  Ciafaloni:2013hya, Bringmann:2013oja}. Therefore we only summarize the main ideas and describe the \ff approach to describe electroweak gauge boson emissions.

\subsection{The fragmentation function approach}

We assume heavy dark matter \textit{i.e.}\ $M_{\rm DM} \gg M_{\rm EW}$, which annihilates into electron-positron pairs only.  The radiation of $W$ and $Z$ bosons off the high energy final-state leptons
is enhanced by logarithms of the form $\ln (M^2_{\rm DM}/M^2_{\rm EW})$ and $\ln^2 (M^2_{\rm DM}/M^2_{\rm EW})$ corresponding to collinear and soft/collinear emission, respectively. No such enhancement
exists for radiation off the non-relativistic initial-state \DM particles, or radiation off intermediate particles. Thus, in the limit $M_{\rm DM} \gg M_{\rm EW}$ the electroweak gauge boson emission is dominated by soft and collinear final-state radiation, which is independent of the specific \DM model or the form of the annihilation cross section.

\begin{figure}[h!]
  \centering
  \includegraphics[width=1\textwidth]{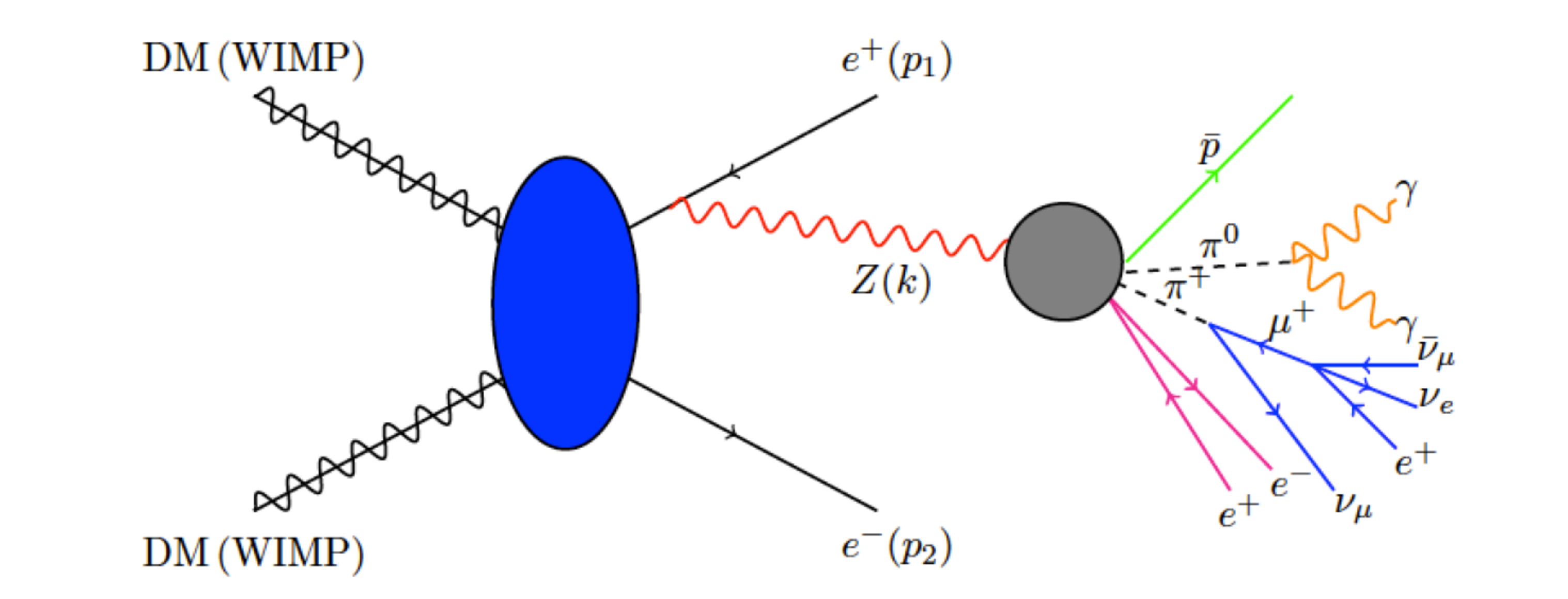}
  \caption{Generic annihilation process of DM into an electron-positron pair plus $Z$ radiation, with $Z$ decay, fragmentation and hadronisation.}
\label{fig0}
\end{figure}

We are interested in the energy spectrum of a given SM final state $f \in \{e^\pm, \gamma, p, \bar{p}, \nu, \bar{\nu}\}$, resulting from the annihilation process ${\rm DM}\; {\rm DM} \to e^+e^- + (Z\to f)$,
including the decay of the $Z$ boson and the fragmentation and hadronisation of the decay products, see Fig~\ref{fig0}.\footnote{For the sake of simplicity, we will
focus only on the radiation of $Z$ bosons in the following.}

The energy spectrum is given by
\begin{equation}\label{eq:dndz}
\frac{dN_f}{dx} = \frac{1}{\langle \sigma v_\text{cm} \rangle }\frac{d\langle \sigma v_\text{cm} \rangle }{dx}\,,
\end{equation}
with $x = 2 E_f/\sqrt{s}$ and $\langle \sigma v_\text{cm}\rangle$ the thermally averaged cross section for the process ${\rm DM}\; {\rm DM}\to e^+e^- (Z \to f)$. The centre-of-mass energy is
$\sqrt{s} = 2 M_\text{DM} / \sqrt{1-v^2_\text{cm}}$, and $E_f$ denotes the energy of the final state SM particle of type $f$.
By $v_\text{cm}$ we denote the velocity of the \DM particle in the centre-of-mass frame.
More specifically, as the \DM particles are non-relativistic, we have $x \simeq E_f/M_{\rm{DM}}$.
In contrast to some conventions in the literature, the energy spectrum, Eq.~(\ref{eq:dndz}), is normalised to one, $\int \! dx \, dN_f/dx = 1$.

The logarithmically enhanced contributions
to the emission of an electroweak boson can be described by fragmentation functions, $D^{\rm{EW}}$, that obey evolution equations
similar to the DGLAP equations \cite{Ciafaloni:2001mu, Ciafaloni:2005fm}. Within this formalism, the energy
spectrum of a SM particle $f$ is given by~\cite{Ciafaloni:2010ti}
\begin{equation}\label{eq:2} \frac{d N_f}{d \ln x}(M_{\rm{DM}},x) = \sum_J \int^1_x dz\, D^{\rm{EW}}_{I\to J}(z) \frac{d N_{J\to f}}{d \ln x}
  \left(zM_{\rm{DM}},\frac{x}{z}\right),
\end{equation}
where $I$ denotes the final state of the $(2\to 2)$ annihilation process, \textit{i.e.}\ an electron-positron pair in the case at hand. The summation in $J$ includes the final state
particles after the radiation of an EW gauge boson. In our case, $J \in \{e^+,e^-,Z\}$. The fragmentation functions $D^{\rm{EW}}_{I\to J}(z)$ describe the probability that a
particle $I$ turns into
a particle $J$, with a fraction $z$ of the energy of the emitting particle, via the emission of a EW boson. Clearly at
the lowest order we have $D^{\rm{EW}}_{I\to J}(z) = \delta_{IJ}\delta(1-z)$. The further decay, fragmentation and hadronisation of the primary annihilation products, $J\to f$, can be described by
standard Monte Carlo event generators.

The fragmentation functions $D^{\rm{EW}}_{I\to J}(z)$ can be obtained by computing the partonic splitting functions. They differ from those of QCD or QED
as the emitted EW gauge bosons are massive.\footnote{For a more extended discussion and a complete listing of the EW splitting functions see Ref.~\cite{Ciafaloni:2010ti}.}
For our purposes, we need the splitting functions of a massless fermion (F) splitting into a massless fermion and a massive vector particle (V):
	\begin{subequations}\label{eq:F1}
		\begin{align}
		P_{F\to F} &= \frac{1+x^2}{1-x} L(1-x), \\
		P_{F\to V} &= \frac{1+(1-x)^2}{x} L(x),
		\end{align}
	\end{subequations}
with
	\begin{equation}\label{eq:F2}
	L(x) = \ln\frac{s x^2}{4 m_Z^2} + 2\ln\left(1+\sqrt{1-\frac{4 m_Z^2}{s x^2}}\right) .
	\end{equation}
The corresponding fragmentation functions are
\begin{subequations}\label{eq:D}
		\begin{align}
		D_{F\to F} &= \frac{\alpha_2}{2\pi \cos^2\theta_w}g^2_f P_{F\to F}, \\
		D_{F\to V} &= \frac{\alpha_2}{2\pi \cos^2\theta_w}g^2_f P_{F\to V},
		\end{align}
\end{subequations}
where $m_Z$ is the mass of the emitted $Z$ boson, $\alpha_2$ the SU(2) coupling and $\theta_w$ the weak angle. The
factor $g_f = T_3^f -\sin^2\theta_w Q_f$ accounts for the coupling of the fermions to the $Z$ boson, where $T_3^f$ and $Q_f$ are the isospin and the charge of the fermion, in our
case $e^\pm_L$ or $e^\pm_R$.
The function $L$, Eq.~(\ref{eq:F2}), respects the correct kinematic limits and vanishes below threshold for $x < 2 m_Z /\sqrt{s}$.  Note that the fragmentation functions depend on the
chirality of the fermion through the coupling $g_f$, so that one should sum over $e^+_Le^-_L$ and $e^+_Re^-_R$ pairs to obtain the annihilation cross section.

For the implementation of the \ff approach in our Monte Carlo program, we use a Sudakov parametrisation of the phase space~\cite{Ciafaloni:2010ti} to produce events according to the \ff distributions.
This is required only because we use a parton shower program to evolve the final state and it needs four momenta as input.
Note that for observables at the annihilation point such as cross sections or distributions, like those in Figs.~\ref{fig3} and \ref{fig6}, Eq.~(\ref{eq:2}) would be enough
to determine the quantities of interest.
The four-momentum of the initial state is denoted by
\eq{ S^{\mu} = \left( 2 E, 0, 0, 0 \right) ,}
where $E = M_{\rm DM} / \sqrt{1- v^2_{\text{cm}}}$. The momentum of the electron/positron that radiates the $Z$ boson can be
parametrised as
\eq{p_1 = \left( E \left( 1-x + \frac{k_t^2}{4 E^2 \left( 1-x \right)} \right), - k_t, 0, E \left( 1-x - \frac{k_t^2}{4 E^2 \left( 1-x \right)} \right) \right), \; \text{with} \; k_t \ll E,}
where $x$ corresponds to the fraction of energy carried away by the emitted $Z$ boson. The four-momentum of the emitted
$Z$ boson takes the form
\eq{ k_Z = \left( E \left( x + \frac{k_t^2 + m^2_Z}{4 E^2 x} \right), k_t, 0, E \left( x - \frac{k_t^2 + m^2_Z}{4 E^2 x}  \right)\right) .}
Finally, the four-momentum of the electron/positron which does not radiate is
\eq{p_2 = \left(  E \left( 1- R(1-x,k_t) \right) , 0, 0, - E \left( 1- R(1-x,k_t) \right) \right) ,}
with
\eq{R(x,k_t) = \frac{k_t^2}{4 E^2 x} + \frac{k_t^2 + m^2_{Z}}{4 E^2 (1-x)} .}
This parametrisation ensures the conservation of the four-momentum.

\section{Majorana fermion and vector dark matter}
\label{ch:models_setup}

In order to quantify the accuracy of the \ff approximation, we have calculated \DM annihilation into an $e^+e^-$-pair plus a $Z$ boson, ${\rm DM}\; {\rm DM}\to e^+e^-Z$, in the minimal supersymmetric 
model (MSSM) with neutralino dark matter and in a model with universal extra dimensions (UED), where the dark matter is provided by the first \KK excitation of the photon. For both models, we have 
compared the energy spectrum of the $Z$ boson within the \ff approximation with the exact calculation of the $(2\to3)$ process, including the radiation of a $Z$ boson from the intermediate $t$- and $u$-
channel particles.
In both cases, we have chosen the intermediate particle to be
  degenerate in mass with the dark matter particle. This choice, on
  the one hand, 
  reduces the number of model parameters. On the other hand, because of an on-shell
  enhancement of the intermediate propagator  for small
  electron/positron energies, the case of equal masses results in a 
  spectrum which is most strongly peaked towards the maximal $Z$-boson energy. 
A priory, one would
thus expect that the \ff approach is less likely to reproduce the full
calculation.
We have used \textsc{FeynArts} and \textsc{FormCalc}~\cite{Hahn:2000kx,Hahn:2001rv,Hahn:1998yk} for the MSSM calculation and \textsc{CalcHep}~\cite{Belyaev:2012qa,Datta:2010us} for 
UED, and checked our results against \textsc{MadGraph~5}~\cite{Alwall:2011uj}.

The flux of the annihilation products is determined by the thermally averaged cross section, $\langle \sigma v_{\rm cm} \rangle $, which can be expanded in powers of the  \DM velocity~\cite
{Srednicki:1988ce}:
\begin{equation}
\langle \sigma v_{\rm cm} \rangle = a + b v^2_{\rm cm} + \mathcal{O}(v^4_{\rm cm}).
\end{equation}
Given that $v^2_{\rm cm} \approx 10^{-6}$ for the annihilation of dark matter in the halo,
we only kept the first term of the expansion.

\subsection{A supersymmetric model with Majorana fermion dark matter}
\begin{figure}[h!]
  \centering
  \includegraphics[width=1\textwidth]{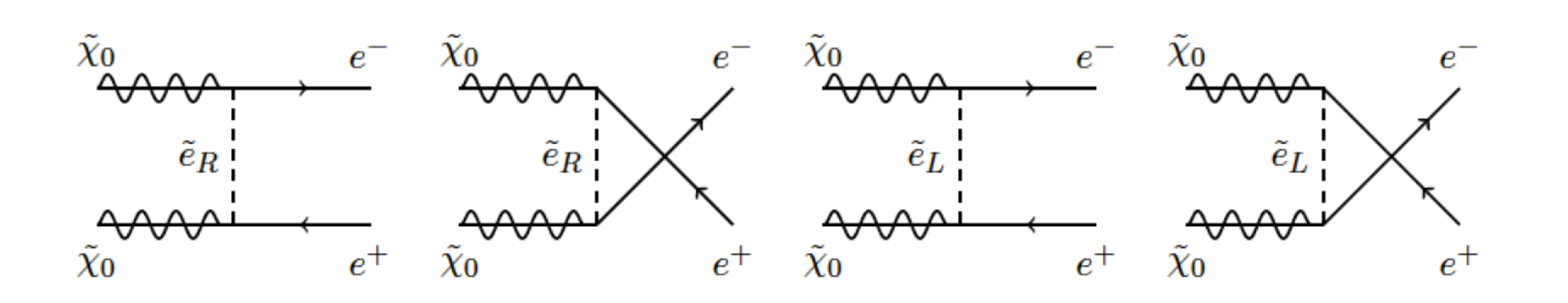}
  \caption{Lowest order contributions to neutralino annihilation into an electron-positron pair.}
\label{fig1}
\end{figure}

As a specific model with Majorana fermion dark matter we have considered the MSSM and calculated the annihilation of neutralinos into an electron-positron pair,
$\tilde{\chi}_0\tilde{\chi}_0 \to e^+ e^-$.  We consider a pure bino, which does not couple to a $Z$ boson, so that the annihilation proceeds
through the exchange of selectrons in the $t$- and $u$-channel only. To derive analytic results, we have furthermore assumed that there is no neutralino and selectron mixing, and that the masses of the 
left- and right-chiral selectrons are degenerate. Without mixing, the vertices drastically simplify to $- i e
\sqrt{2} P_L /{(2 \cos \theta_w)}$ for the left-chiral selectron, $\tilde{e}_L$, and $i e \sqrt{2} P_R/\cos \theta_w$ for
the right-chiral selectron, $\tilde{e}_R$, respectively~\cite{Haber:1984rc, DrRoGo04}. The Feynman diagrams for this process are displayed in Fig.~\ref{fig1}.
We have reproduced the result for the matrix element squared obtained in~\cite{Griest:1988ma} and performed a low-velocity expansion of the
Mandelstam variables to obtain the thermally averaged cross section~\cite{We94,Roszkowski:1994tm}. Retaining only the first coefficient of this expansion {\it i.e.}\ the zeroth order in
the velocity, we obtain
\eqa{\label{eq:susy}\langle \sigma v_{\rm cm} \rangle = \frac{\alpha^2 \,25 \pi\, m_e^2 \sqrt{M_{\rm DM}^2-m_e^2}}{ 16 M_{\rm DM} \cos^4 \theta_w \left(M_{\rm DM}^2+m^2_{\tilde{e}}-m_e^2\right)^2} ,}
where $m_e$ and $m_{\tilde e}=m_{\tilde{e}_L}=m_{\tilde{e}_R}$ are the mass of the electron and of the selectrons, respectively, and $\alpha$ is the QED
coupling. In the following we will always set $m_{\tilde e} = M_{\rm DM}$ for simplicity. Note that because of the well-known helicity
suppression~\cite{Goldberg:1983nd,Go832} due to the Majorana nature of the neutralino the cross section vanishes in the limit of zero electron mass $m_e \to 0$.

\begin{figure}[h!]
  \centering
  \includegraphics[width=0.49\textwidth]{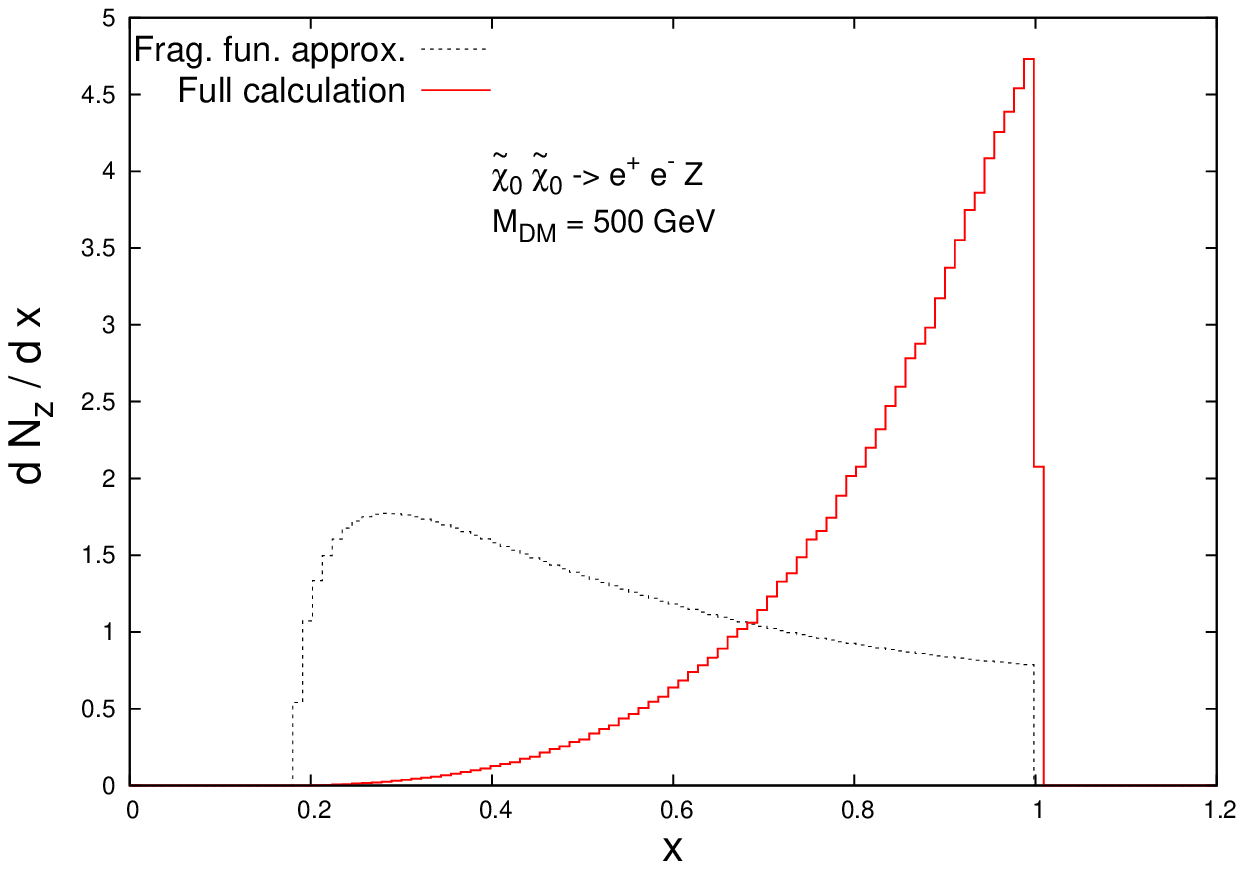}
  \includegraphics[width=0.49\textwidth]{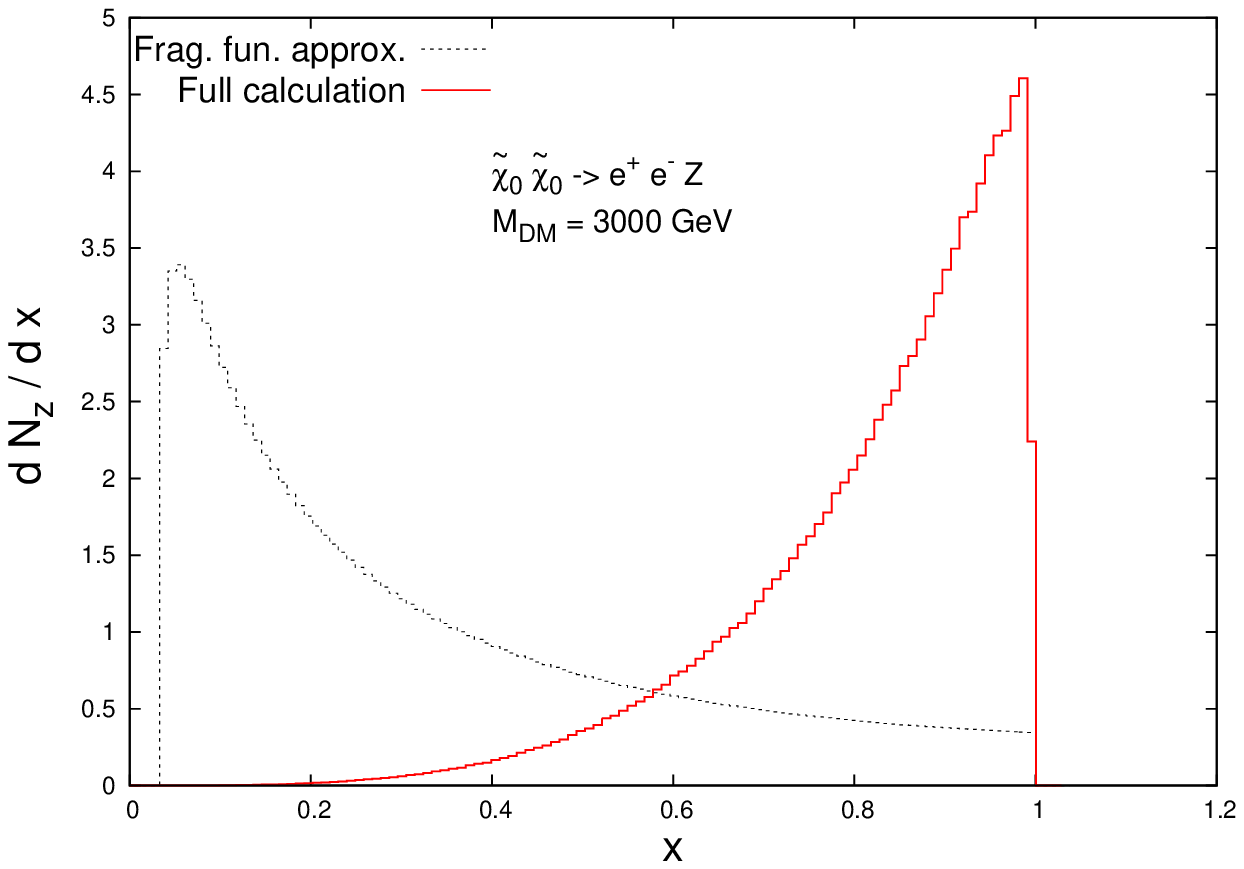}
  \caption{$Z$-boson energy spectrum $dN_Z/dx$, Eq.~(\ref{eq:dndz}), for the annihilation of Majorana fermion dark matter into an electron-positron pair plus a $Z$ boson in supersymmetry. (See the main 
text for the definition of the model.) Shown is the exact calculation (red, solid line) and the \ff approximation (blue, dotted line) for  $M_{\rm DM} = 500$ and 3000 GeV.}
\label{fig3}
\end{figure}

The $(2 \to 3)$ tree-level process, $\tilde{\chi}_0\tilde{\chi}_0 \to e^+ e^-Z$, is finite, and we do not include virtual corrections as we are mainly interested in the shape of the secondary flux 
induced by the $Z$-boson decay. The helicity suppression of the ($2\to 2$) process, Eq.\,(\ref{eq:susy}), is lifted by $Z$-boson radiation only if the emission from both the final- and intermediate 
state particles, \textit{i.e.} from the electron-positron pair and from the $t$- and $u$-channel selectron, is included~\cite{Ciafaloni:2011sa}. The \ff approximation which only includes soft/collinear 
radiation off the electron/positron pair is thus expected not to work in this particular case.
This can be seen explicitly from the s-wave contribution to the neutralino annihilation cross section, $\tilde{\chi}_0\tilde{\chi}_0 \to e^+ e^-Z$, obtained in Ref.~\cite{Ciafaloni:2011sa}:
\begin{equation}
\frac{d N_Z}{dx} =  \frac{\left( 1-x \right)}{\left( 2-x \right)^2} \left[ \frac{\left( 1-x \right)}{\left( 2-x \right)} \ln \left( 1-x \right) + x \frac{\left( 1-x \right)^2 + 1 }{4 \left( 1-x 
\right)} \right],
\end{equation}
where $x$ is the energy fraction of the emitted $Z$ boson and the mass of the electron is assumed to be zero.
In Fig.~\ref{fig3} we compare the $Z$ energy distribution of the full $(2\to3)$ calculation for bino annihilation into an electron-positron pair plus a $Z$ boson with the \ff approximation
\begin{equation}\label{eq:splitting_sigma}
\left.\frac{d\sigma_{(\mathrm{DM}\,\mathrm{DM}\to e^+ e^- Z)}}{dx}\right|_\text{ff} = 2
\left(\sigma_{(\mathrm{DM}\,\mathrm{DM}\to e^+_L e^-_L)}D_{e_L\to Z}+\sigma_{(\mathrm{DM}\,\mathrm{DM}\to e^+_R e^-_R)}D_{e_R\to Z}\right).
\end{equation}
It is manifest from Fig.~\ref{fig3} that the \ff approximation does not describe the energy distribution of the full $(2\to 3)$ process.
In general, the \ff approach is not supposed to work if the lowest-order cross section is helicity suppressed, as in the case of Majorana fermion annihilation into light fermions.
Moreover, since the splitting functions are convoluted with the $2 \to 2$ cross section, the $2 \to 3$ cross section obtained from the \ff approach would be zero in the case of massless electron/
positron.

\subsection{A universal extra dimension model with vector dark matter}

\begin{figure}[h!]
  \centering
  \includegraphics[width=1\textwidth]{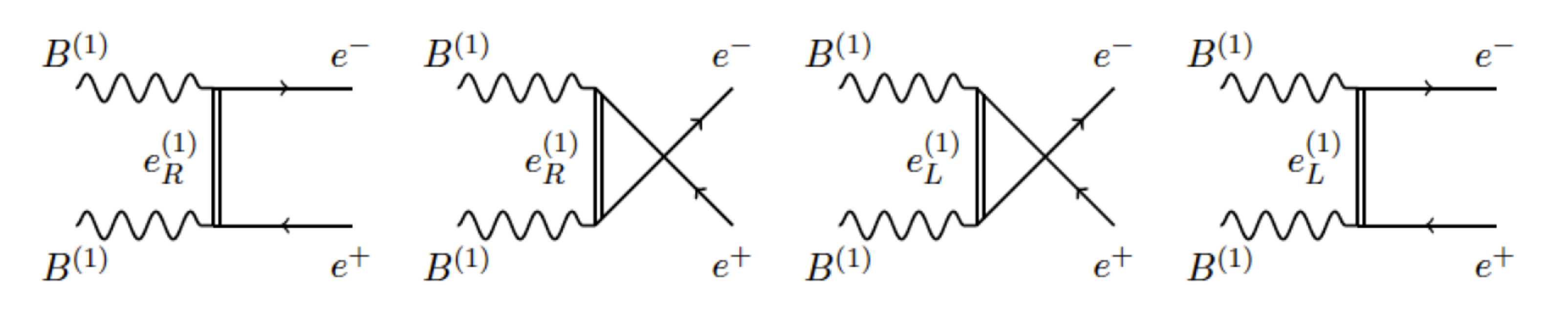}
  \caption{Lowest order contributions to \KK photon annihilation into an electron-positron pair.}
\label{fig10}
\end{figure}

We now discuss a model with vector dark matter, where the lowest-order annihilation cross section to electron-positron pairs is not helicity suppressed. Specifically, we consider a model with universal 
extra dimensions where the first \KK excitation of the U(1)$_Y$ hypercharge gauge field, $B^{(1)}$, provides a dark matter candidate. The Feynman diagrams for the leading-order annihilation process,
$B^{(1)}B^{(1)} \to e^+e^-$, mediated by $t$- and $u$-channel exchange of the first \KK excitation of the electron, $e_{L,R}^{(1)}$, are displayed in Fig.~\ref{fig10}.
For completeness we give the $s$-wave contribution of the thermally averaged annihilation cross section at leading-order~\cite{Servant:2002aq,Kong:2005hn}:
\eq{\label{eq:UED_sigma_2to2}
\langle \sigma v_{\rm cm} \rangle = \frac{\left(|g_L|^4 + |g_R|^4\right)}{576 M_\text{DM}^2 \pi} \,,}
where $g_{L/R} = g_1 Y_{L/R}$ and the hypercharges for left- and right-handed electrons are $Y_L = - 1$ and $Y_R = - 2$, respectively.
For simplicity, and to derive compact analytic expressions, we assume that all \KK particles are mass degenerate.
From Eq.~\eqref{eq:UED_sigma_2to2} it is clear that the process $B^{(1)}B^{(1)} \to e^+e^-$ is not helicity suppressed.

The emission of a $Z$ boson in the $(2\to 3)$ process $B^{(1)}B^{(1)} \rightarrow e^+e^-Z$ constitutes a genuine
higher-order correction of ${\cal O}(\alpha \ln^2 (M^2_{\rm DM}/m^2_Z))$, and is thus significantly suppressed with respect to the lowest order cross section presented in Eq.~(\ref{eq:UED_sigma_2to2}), see Table~\ref{Tablevalues}.
\begin{table}[h!]
\renewcommand{\arraystretch}{1.2}
\centering
\begin{tabular}{|c|c|c|c|}
\hline
  {\small $M_{\rm DM}$ [GeV]}
 & {\small $\langle\sigma v\rangle_{B^{(1)}B^{(1)} \rightarrow e^+e^-}$ [pb]}
 & {\small $\langle\sigma v\rangle_{B^{(1)}B^{(1)} \rightarrow e^+e^-Z}$ [pb]}
 & {\small $\langle\sigma v\rangle_{B^{(1)}B^{(1)} \rightarrow e^+e^-Z}|_{\rm ff}$ [pb]}\\
\hline
  150
 & $ 2.642 $
 & $ 4.583 \times 10^{-3}$
 & $ 1.459 \times 10^{-3}$ \\
\hline
  300
 & $ 0.6604 $
 & $ 2.078 \times 10^{-3}$
 & $ 1.597 \times 10^{-3}$\\
\hline
  500
 & $ 0.2378 $
 & $ 1.202 \times 10^{-3}$
 & $ 1.104 \times 10^{-3}$\\
\hline
  1000
 & $ 5.944 \times 10^{-2} $
 & $ 5.362 \times 10^{-4}$
 & $ 5.282 \times 10^{-4}$\\
\hline
  3000
 & $ 6.605 \times 10^{-3} $
 & $ 1.221 \times 10^{-4} $
 & $ 1.227 \times 10^{-4} $\\
\hline
\end{tabular}
\caption{Thermally averaged cross section $\langle\sigma v\rangle$ for the annihilation of vector dark matter into an electron-positron pair, and with the radiation of a $Z$ boson, in a
universal extra dimension model. (See the main text for the definition of the model.). Shown are the lowest-order cross section, the exact cross section including $Z$-boson radiation, and the \ff 
approximation to $Z$-boson radiation, for different masses of the dark matter particle. We have used $\sin^2 \theta_w = 0.23113$ and $\alpha = 1/128$.}
\label{Tablevalues}
\end{table}
The prediction for the $(2\to 3)$ process $B^{(1)}B^{(1)} \rightarrow e^+e^-Z$ in the \ff approximation includes the logarithmically enhanced contributions $\propto \ln^{(1,2)} (M^2_{\rm DM}/m^2_Z))$ 
only, and is thus expected to become more and more accurate with increasing dark matter mass.
This is born out by the explicit numerical comparison shown in
Table~\ref{Tablevalues}.
While the \ff approximation underestimates the cross section by about a factor 3 for $M_{\rm DM} = 150$\,GeV, at
$M_{\rm DM} = 500$\,GeV it already reproduces the normalisation of the full calculation within 10\% accuracy.

\begin{figure}[h!]
  \centering
  \includegraphics[width=0.49\textwidth]{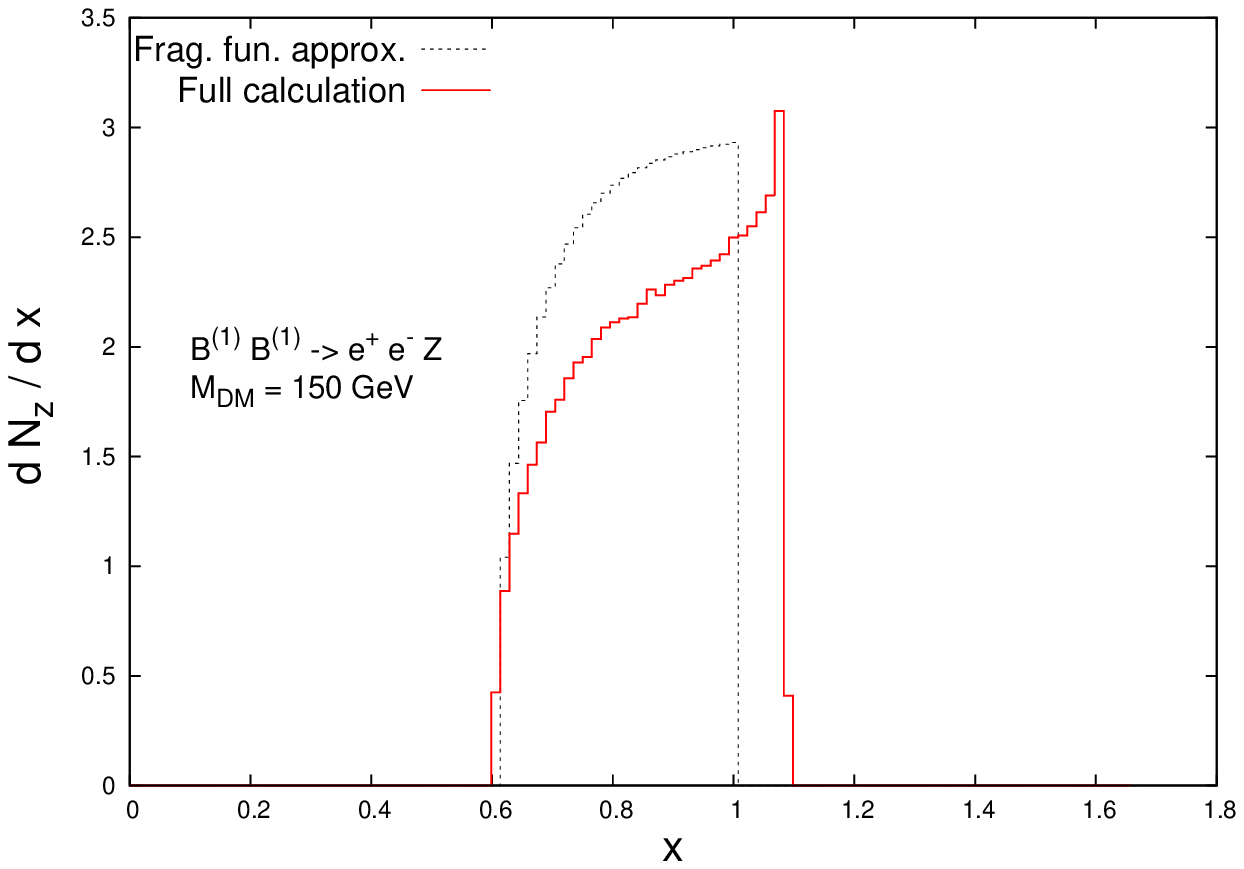}
  \includegraphics[width=0.49\textwidth]{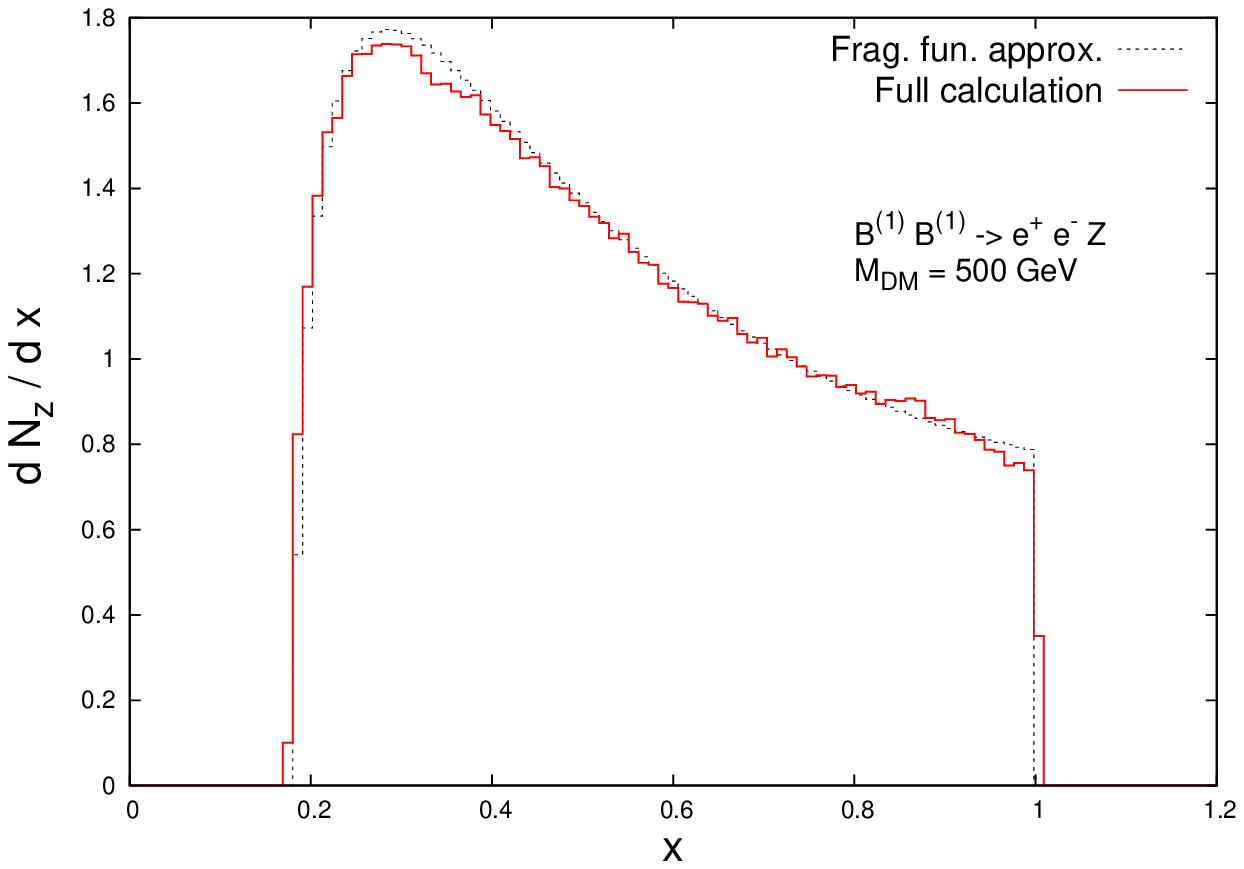}
  \includegraphics[width=0.49\textwidth]{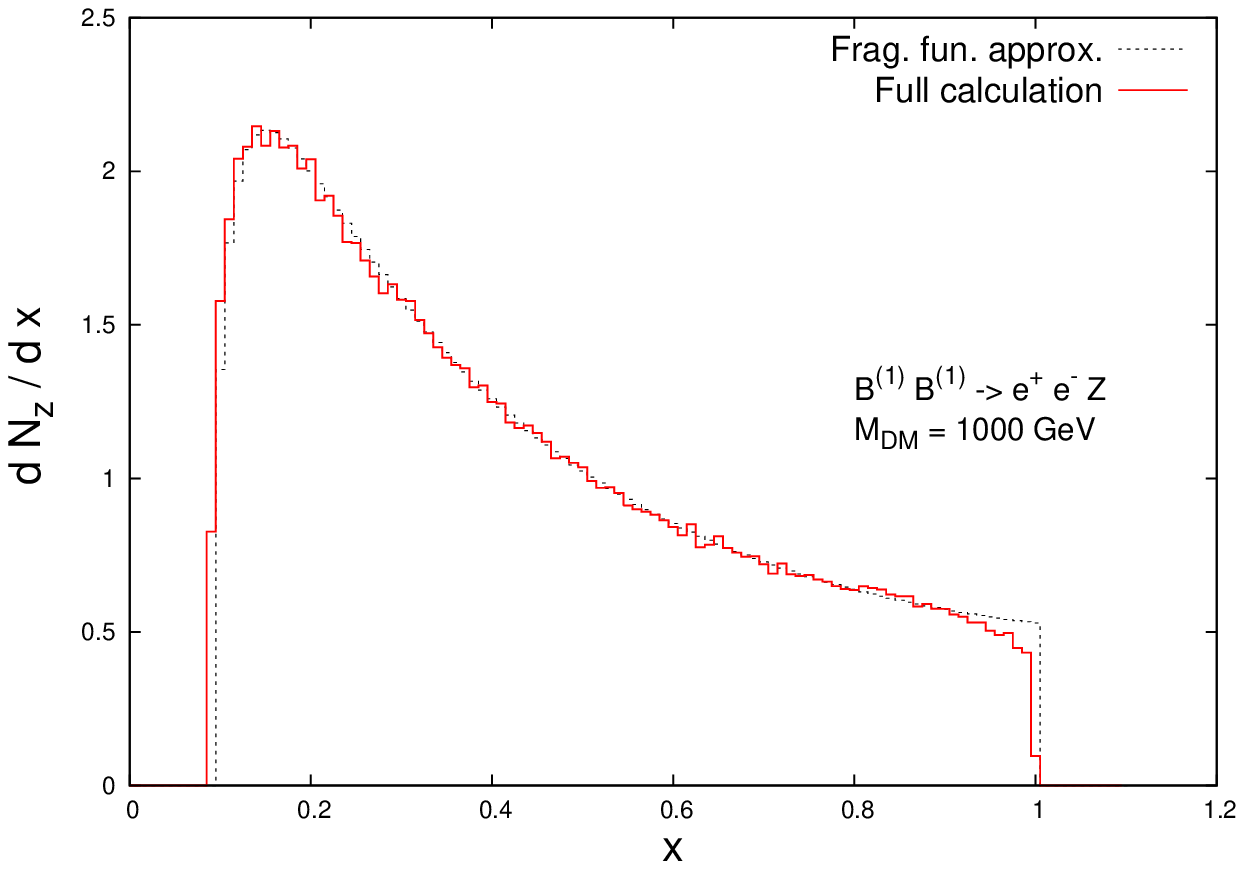}
  \includegraphics[width=0.49\textwidth]{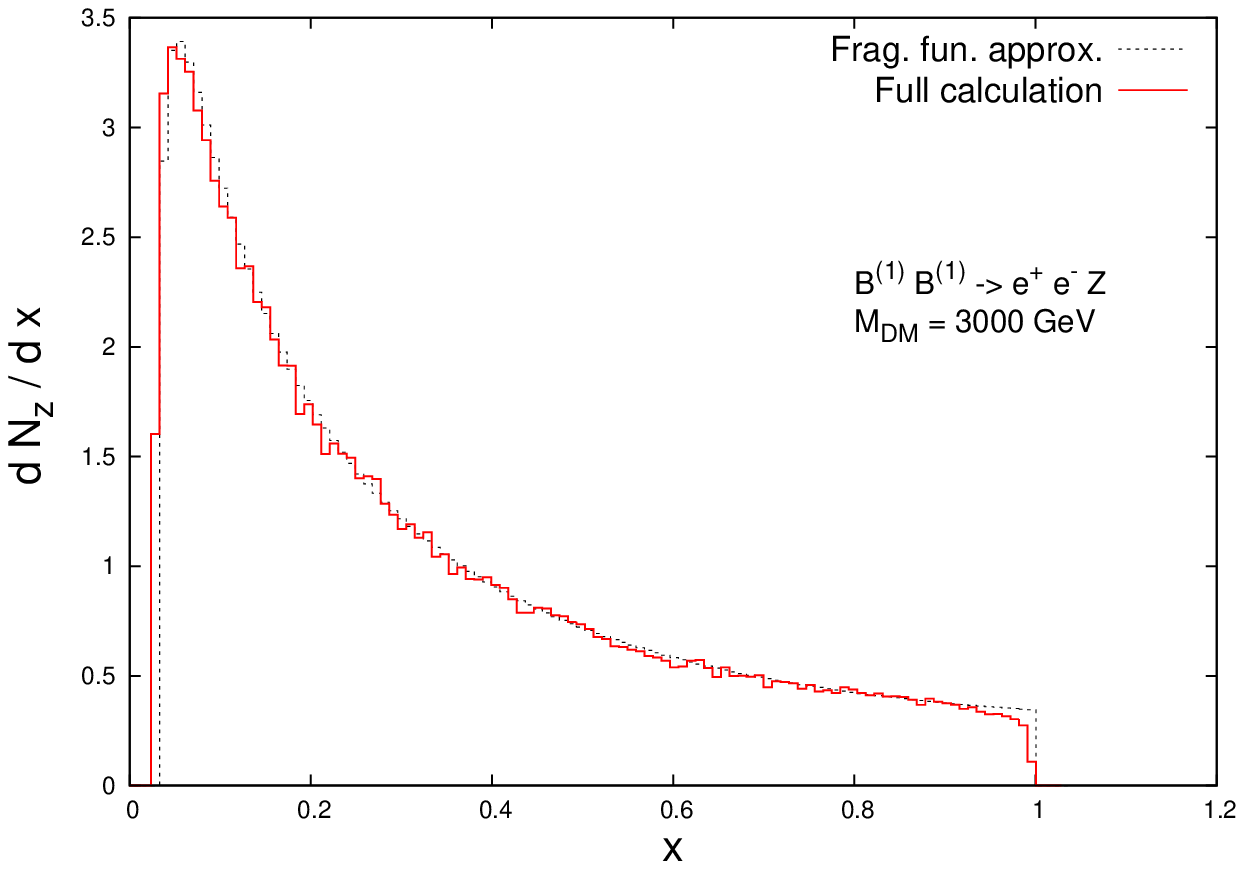}
  \caption{$Z$-boson energy spectrum $dN_Z/dx$, Eq.~(\ref{eq:dndz}), for the annihilation of vector dark matter into an electron-positron pair plus a $Z$ boson in a universal extra dimension model. (See the main text for the definition of the model.) Shown is the exact calculation (red, solid line) and the \ff approximation (blue, dotted line) for  $M_{\rm DM} = 150, 500, 1000$ and 3000 GeV.}
\label{fig6}
\end{figure}

The energy distributions of the emitted $Z$ boson, $dN_Z/dx$, is shown in Fig.~\ref{fig6} for different masses of the dark matter particle.
One can see that the \ff approximation provides a very accurate description of the shape of the energy distributions for heavy dark matter
with $M_{\rm DM}/m_Z \gtrsim 5$.

The quality of the \ff approximation can be deduced from the analytical expression that we have obtained for the exact $(2\to 3)$ process
$B^{(1)}B^{(1)} \rightarrow e^+_L e^-_L Z$.
Integrating over the angles and considering the centre of mass of the annihilating particles, the differential cross section can be written as

\eq{\label{labeleq2} v d \sigma = \frac{|\mathcal{M}|^2}{256 \pi^3} dx_1 dx_2 ,}
where $|\mathcal{M}|^2$ is the matrix element squared and $x_1$ and $x_2$ parametrise the energies of the final-state particles.
In particular
\eqa{k^0 &=& (1-x_2) \sqrt{s} / 2 , \\
p_1^0 &=& x_1 \sqrt{s} / 2 , \\
p_2^0 &=& (1-x_1 + x_2) \sqrt{s} / 2 ,}
with the four momenta $k, p_1$ and $p_2$ corresponding to the $Z$ boson, the positron and the electron, respectively.
The integration limits of the phase space for the variable $x_1$ are

\eq{x_- \leq x_1 \leq x_+ ,}
with
\eq{x_{\pm} = \frac{1+x_2}{2} \pm \sqrt{\frac{(1-x_2)^2}{4} - \frac{m_Z^2}{s}} . }
For $x_2$ the integration boundaries are

\eq{-\frac{m_Z^2}{s} \leq x_2 \leq 1 - 2 \frac{m_Z}{\sqrt{s}} .}
Expanding in the velocity and integrating over $x_1$, Eq. \eqref{labeleq2} becomes

\eq{\label{labeleq} \frac{d\langle\sigma v\rangle}{d x_2} = \frac{\alpha}{2304\, M_\text{DM}^2\pi^2} \frac{(1-2 \sin^2 \theta_w)^2}{\sin^2 \theta_w \cos^2 \theta_w} |g_L|^4 F(x_2) ,}
The function $F$ is defined as

\eqa{\label{Ffcuntion} F(x_2) &=& \Bigg( A \ln\left( \frac{\overline{x}_+}{\overline{x}_-}\right) + B +  C \ln \left( \frac{\overline{x}_+ -2}{\overline{x}_- - 2} \right) \Bigg) ,}
with $\overline{x}_{\pm} = 1 - x_2 \pm \sqrt{(1-x_2)^2 - m_Z^2/M_\text{DM}^2}$ and coefficients $A$, $B$ and $C$ given by
\eqa{A &=& - \frac{1}{8 M^6_{\text{DM}}} \frac{1}{x_2 (x_2 - 1)} \Big( m_Z^6 + M^2_{\text{DM}} m_Z^4 (4+5 x_2) \nonumber \\
&& + 4 M^4_{\text{DM}} m_Z^2 (1 + 3 x_2 + 2x_2^2) + 8 M^6_{\text{DM}} (x_2 + x_2^3) \Big)  , \nonumber \\
B &=& - \sqrt{(1-x_2)^2-\frac{m^2_Z}{M^2_{\text{DM}}}} + \frac{1}{16 M^6_{\text{DM}}} \frac{1}{(1+x_2)^2}\sqrt{(1-x_2)^2-\frac{m^2_Z}{M^2_{\text{DM}}}} \frac{1}{x_2 + \frac{m_Z^2}{4 M^2_{\text{DM}}}} \Big(2 m_Z^6 \nonumber \\
&& + 2 M^2_{\text{DM}} m_Z^4 (2+9x_2) + 12 M^6_{\text{DM}} x_2 (5 + 8 x_2 + 5x_2^2) + 5 M^4_{\text{DM}} m_Z^2 (3+8x_2 + 11 x_2^2) \Big) , \nonumber \\
C &=& \frac{1}{8 M_{\text{DM}}^6} \Big( M^2_{\text{DM}} m^4_Z (4 + 9 x_2 - 4 x_2^2 + 5x_2^3) + 8 M_{\text{DM}}^6 x_2 (1+4x_2 + 9x_2^2 + 4x_2^3 + x_2^4) \nonumber \\
&& + m^6_Z (1+x^2_2) + M_{\text{DM}}^4 m_Z^2 (4 + 25 x_2 + 36 x_2^2 - 7 x_2^3 + 8 x_2^4) \Big) \frac{1}{x_2 (1+x_2)^3} .}
Given the Born cross section (see Eq. \eqref{eq:UED_sigma_2to2}),  Eq. \eqref{labeleq} can be recast in the form

\eq{ \frac{d \langle\sigma v\rangle}{d x_2} = 2\langle\sigma v\rangle_{\text{Born}} \frac{\alpha}{2 \pi} \frac{g_f^2}{\sin^2 \theta_w \cos^2 \theta_w} F(x_2) .}
This is exactly the form expected within the \ff  approach, see Eqs.~\eqref{eq:F1}, \eqref{eq:F2} and \eqref{eq:D}.
Indeed in the limit $m_Z \to 0$,

\eq{ A \ln\left( \frac{\overline{x}_+}{\overline{x}_-}\right) \to \frac{1+x_2^2}{1-x_2} \left( \ln\left( \left(1-x_2\right)^2 \frac{M^2_{\text{DM}}}{m^2_Z} \right) + 2 \ln\left( 2 \right) \right), }
which is exactly Eqs.~\eqref{eq:F1} and \eqref{eq:F2} in this limit.

For completeness we have computed the coefficients $A$, $B$ and $C$ for an intermediate mass $m_i$ different from the \DM mass in the limit $m_Z \to 0$.
In this case they read
\eqa{A &=& \frac{4}{(1+w^2)^2} \frac{-1+ 4 x_2 - 3 x_2^2 +2 x_2^3 + w^2 \left( 1 - 2 x_2 + 3 x_2^2 \right) }{(1-x_2) (-1 + w^2 + 2 x_2)} , \nonumber \\
B &=& -\frac{1+w^2}{2} (1-x_2) + \frac{1}{8 (1+w^2)} \frac{1}{(w^2 + x_2)^2} \frac{(x_2 - 1)}{w^4 - 1 + 2 x_2 (1+w^2)}  \nonumber \\
&& \Bigg( -3 + w^{10} + w^8 \left( -3 + x_2 \right) -14 x_2 - 28 x_2^2 - 20 x_2^3 + w^6 \left( 6 - 23 x_2 + 6 x_2^2 - 2x_2^3 \right) \nonumber \\
&& - w^4 \left( 2 - 7 x_2 + 44 x_2^2 + 2x_2^3 \right) - w^2 \left( -1 + 31 x_2 + 30 x_2^2 + 36 x_2^3 \right) \Bigg) \nonumber \\
C &=& \frac{1}{(w^2 + x_2)^3} \frac{1}{4(1+w^2)^2} \frac{1}{(-1+w^2+2x_2)} \Bigg( 3 - w^{14} - 3 w^{12} (-1 + x_2) + 8 x_2  \\
&& - 10 x_2^2 - 20 x_2^3 - 16 x_2^4 + 32 x_2^5 - w^{10} \left( - 9 -31 x_2 - 6x_2^2 + 2 x_2^3 \right) \nonumber \\
&& - w^2 \left(1-7x_2 + 14 x_2^2 + 110 x_2^3 - 148x_2^4 \right) - w^4 \left( 11 - 21 x_2 + 176 x_2^2 - 308 x_2^2 - 4 x_2^4 \right) \nonumber \\
&& - w^6 \left( -9 +102 x_2 -232 x_2^2 - 96 x_2^3 +4 x_2^4 \right) - w^8 \left( 11 - 70 x_2 - 90 x_2^2 -16 x_2^3 + 4 x_2^4 \right) \Bigg) \nonumber ,} 
with $w= m_i / m_\text{DM}$. Furthermore, in Eq.~\eqref{Ffcuntion} $\ln \left( \frac{\overline{x}_+ -2}{\overline{x}_- - 2} \right)$ has to be replaced by $\ln \left( \frac{\overline{x}_+ - (1+w^2)}{\overline{x}_- - (1+w^2)} \right)$.

\section{The secondary flux after propagation}
\label{ch:pythia}

In this section we compare the positron and antiproton fluxes obtained with the \ff approximation against the exact result
after propagation through the galactic halo. We briefly explain how we obtain the energy spectra of the stable Standard Model particles
at the production point and how we implement the propagation of positrons and antiprotons.

In order to obtain the primary\footnote{``Primary" denotes the particles at the production point, while ``secondary"
  denotes the particles after propagation.} stable SM particles we use \textsc{Pythia~8} \cite{Sjostrand:2006za,Sjostrand:2007gs,Sjostrand:2008vc}
  to describe the $Z$ boson
decay, the evolution of its decay products (further decays and hadronisation) and the QCD and QED radiation. The fluxes produced
by \textsc{Pythia~8} are then propagated through the galactic halo to obtain predictions for positrons and antiprotons at the earth.

As we are not aiming at a detailed study of the different dark matter profiles, and the uncertainty due to propagation, we have chosen the Einasto
model~\cite{Graham:2005xx, Navarro:2008kc} as a specific example:
\eq{\rho_{\text{Ein}} (r) = \rho_s \exp \left( - \frac{2}{\alpha} \left[ \left( \frac{r}{r_s}\right)^{\alpha} - 1  \right] \right) ,}
where $\alpha = 0.17$, $r_s = 28.44 \; \text{kpc}$ and $\rho_s = 0.033 \; \text{GeV} / \text{cm}^3$. These values
correspond to a dark matter density $\rho_{\odot} = 0.3 \;
\text{GeV} / \text{cm}^3$ at the location of the sun ($r_{\odot} = 8.33 \; \text{kpc}$).

In order to calculate the flux of positrons/electrons at the location of the earth, we have used the Green function
formalism \cite{Cirelli:2010xx}. The expression for the positron/electron flux after propagation is~\cite{Cirelli:2010xx}
\eq{ \label{eq:dphide}\frac{d \Phi_{e^{\pm}}}{d E} (\epsilon, r_{\odot}) = \frac{1}{4 \pi} \frac{v_{e^{\pm}}}{b_T(\epsilon)} \frac12 \left( \frac{\rho_{\odot}}{M_{\rm DM}} \right)^2 \langle\sigma v_{\text{cm}}\rangle_{\text{DM DM}\to \text{I}}
\int^{M_{\rm DM}}_{\epsilon} d \epsilon_s \frac{d N_{e^{\pm}}}{d E} (\epsilon_s) \mathcal{I} \left(\lambda_D(\epsilon, \epsilon_s)  \right) ,}
with
\begin{equation}
\frac{d N_{e^{\pm}}}{d E} = \frac{1}{\langle\sigma v_{\text{cm}}\rangle_{\text{DM DM}\to \text{I}}}\frac{d\langle\sigma v_{\text{cm}}\rangle_{\text{DM DM}\to \text{I}}
\times BR_{\text{I}\to e^\pm}}{dE} ,
\end{equation}
with $\epsilon$ the energy expressed in GeV, $I \in \{e^+ e^-, e^+ e^- Z\}$ and $v_{e^{\pm}}$ is the velocity of the
electron.
The energy spectrum is normalized to the total cross section. The normalisation of the flux is given by $b_T(\epsilon) =~\epsilon^2
/ \tau_{\odot}$ where $\tau_{\odot} = \text{GeV} / b\,(1 \; \text{GeV}, r_{\odot}) = 5.7 \times 10^{15} \; \text{sec}$.
The functions $\lambda_D$ and $\mathcal{I}$ are defined as
\eq{\lambda_D = \lambda_D (\epsilon, \epsilon_s) = \sqrt{4 \mathcal{K}_0 \tau_{\odot} (\epsilon^{\delta - 1} -
    \epsilon^{\delta - 1}_s)/(1-\delta)}}
and
\eq{\mathcal{I} (\lambda_D) = a_0 + a_1 \tanh \left( \frac{b_1 - \ell}{c_1}\right) \left[a_2 \exp \left( - \frac{\left(
          \ell - b_2 \right)^2}{c_2} \right) + a_3 \right] ,}
with $\ell = \log_{10}(\lambda_D / \text{kpc})$. The values of the astrophysical parameters $\mathcal{K}_0$, $\delta$,
$a_0$, $a_1$, $a_2$, $a_3$, $b_1$ and $b_2$ are taken from Ref.~\cite{Cirelli:2010xx} and correspond to the
medium (MED) Einasto profile: ($a_0$, $a_1$, $a_2$, $a_3$, $b_1$, $b_2$, $c_1$, $c_2$, $\delta$, $\mathcal{K}_0$) =
(0.507, 0.345, 2.095, 1.469, 0.905, 0.741, 0.160, 0.063, 0.70, 0.0112).

The flux of antiprotons from dark matter annihilation at the earth is given by~\cite{Cirelli:2010xx}
\eq{ \label{eq:dphidk} \frac{d \Phi_{\bar{p}}}{d K} (\epsilon, r_{\odot}) = \frac12 \frac{v_p}{4 \pi} \left( \frac{\rho_{\odot}}{M_{DM}}
  \right)^2 R(K) \langle\sigma v_{\text{cm}}\rangle \frac{d N_{\overline{p}}}{d K} ,}
where $K = E - m_p$ and $m_p$ are the kinetic energy and the mass of the proton respectively. The function $R$ is
defined as
\eq{\log_{10} \left[ R(K) / \text{Myr} \right] = a_0 + a_1 \kappa + a_2 \kappa^2 + a_3 \kappa^3 + a_4 \kappa^4 + a_5 \kappa^5, }
with $\kappa = \log_{10}( K / \, \text{GeV})$ and the coefficients again taken from Ref.~\cite{Cirelli:2010xx}:
($a_0$, $a_1$, $a_2$, $a_3$, $a_4$, $a_5$) = (1.8804, 0.5813, $-0.2960$, $-0.0502$, $0.0271$, $-0.0027$).

\subsection{Comparison and results}
\begin{figure}[h!]
  \centering
  \includegraphics[width=0.49\textwidth]{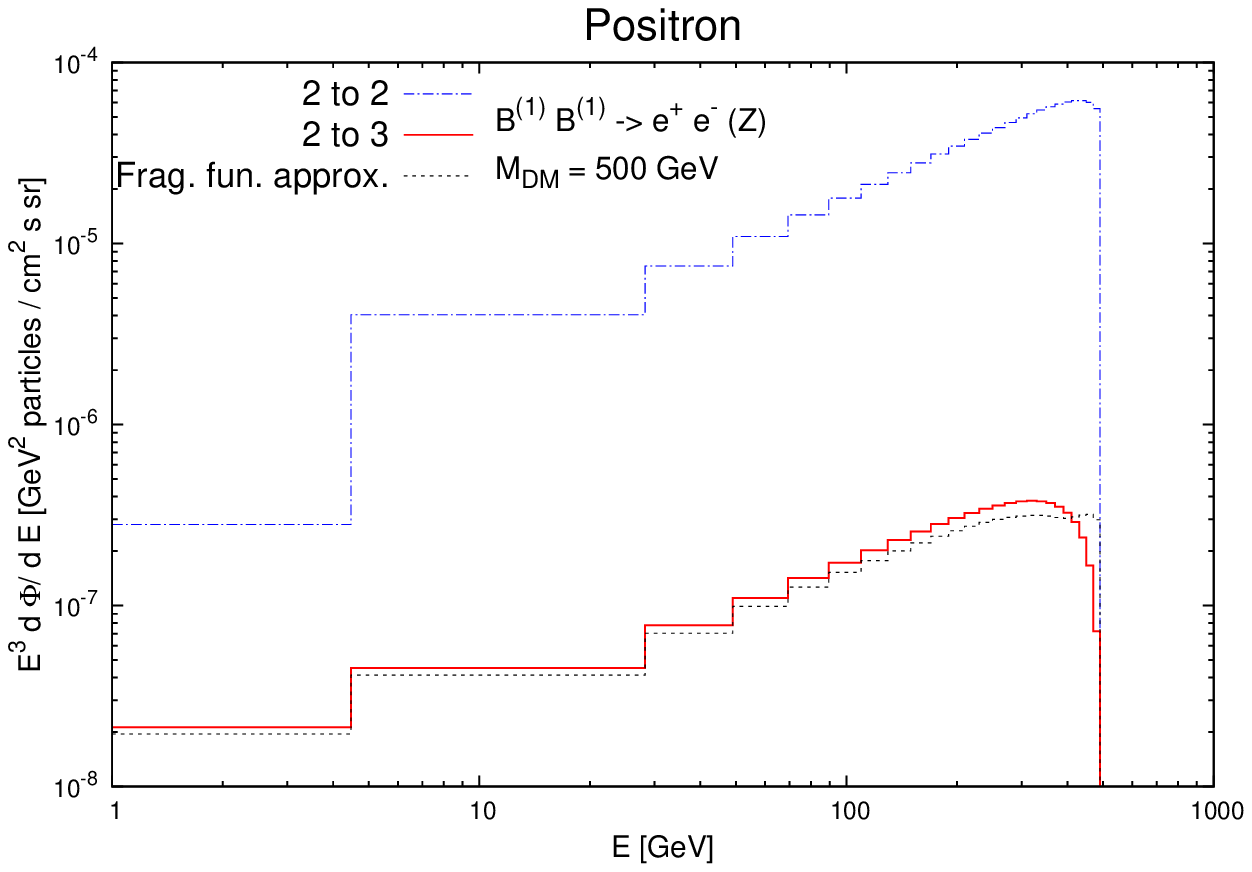}
  \includegraphics[width=0.49\textwidth]{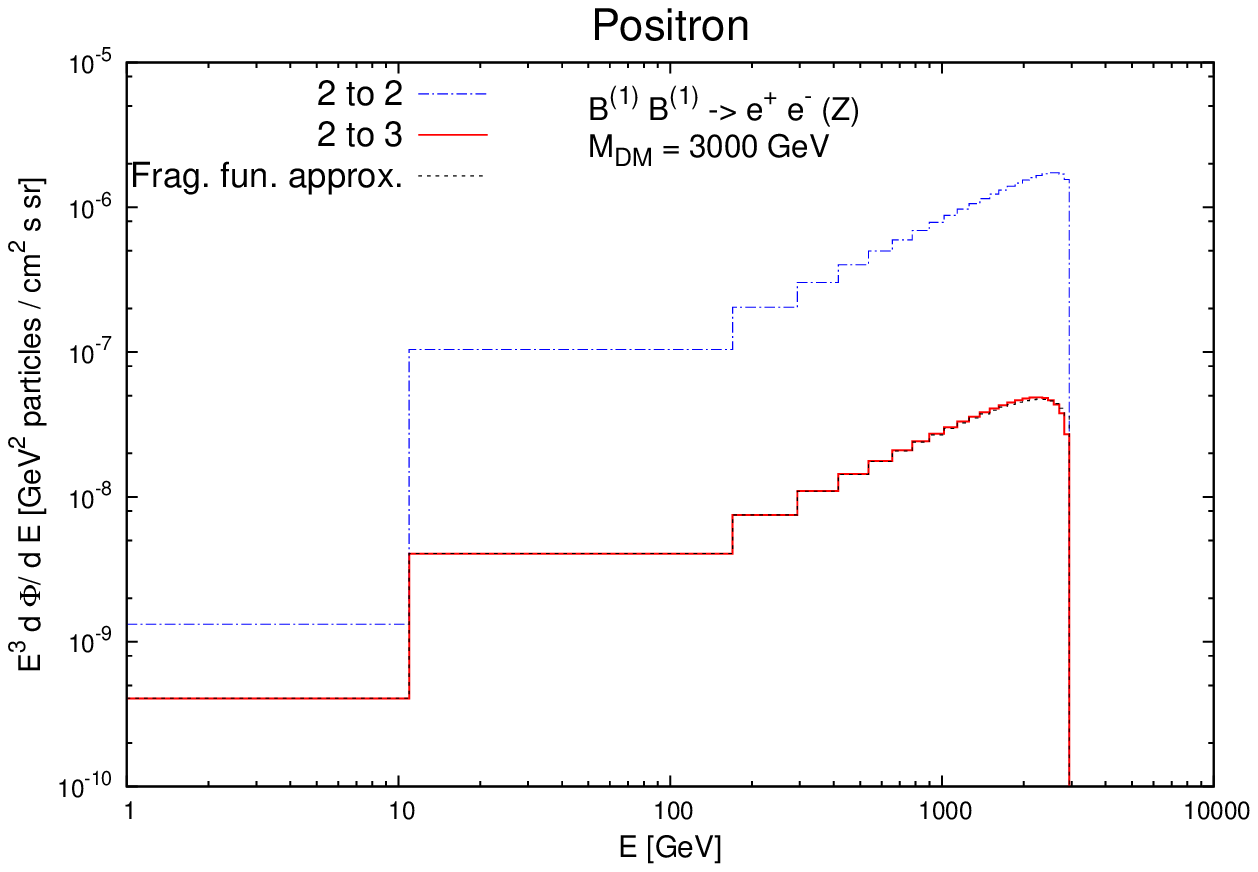}
\caption{Positron flux $d\Phi/dE$, Eq.~(\ref{eq:dphide}), after propagation through the galactic halo for the annihilation of vector dark matter into an electron-positron pair plus a $Z$ boson in a universal extra dimension model. (See the main text for the definition of the model.) Shown is the lowest-order flux (blue, dashed line), the exact calculation of the flux from $Z$ radiation (red, solid line) and the \ff approximation (black, dotted line) for  $M_{\rm DM} = 500$ and 3000 GeV.}
\label{fig8}
\end{figure}

We consider the spectra for positrons and antiprotons from the annihilation of vector dark matter in the universal extra dimension model after propagation through the galactic halo. We compare
the results from the full $(2\to 3)$ calculation and the \ff approximation. In Fig.~\ref{fig8}
the positron spectra after parton shower and propagation are displayed.
The \ff approximation provides an accurate description of the positrons from the $Z$-boson decay. However, as the cross section of the $(2 \to 3)$ process is a genuine electroweak higher-order contribution of ${\cal O}(\alpha \ln^2 (M^2_{\rm DM}/m^2_Z))$, and thus highly suppressed by comparison to the leading-order annihilation, the amount of additional
positrons is small compared to those produced in the $(2\to 2)$ process.
The small dip in the \ff prediction at high energies is a remnant of the kinematics of the
$2 \to 2$ process and is disappearing as $M_{\rm DM}/m_Z$ increases.

\begin{figure}[h!]
  \centering
  \includegraphics[width=0.49\textwidth]{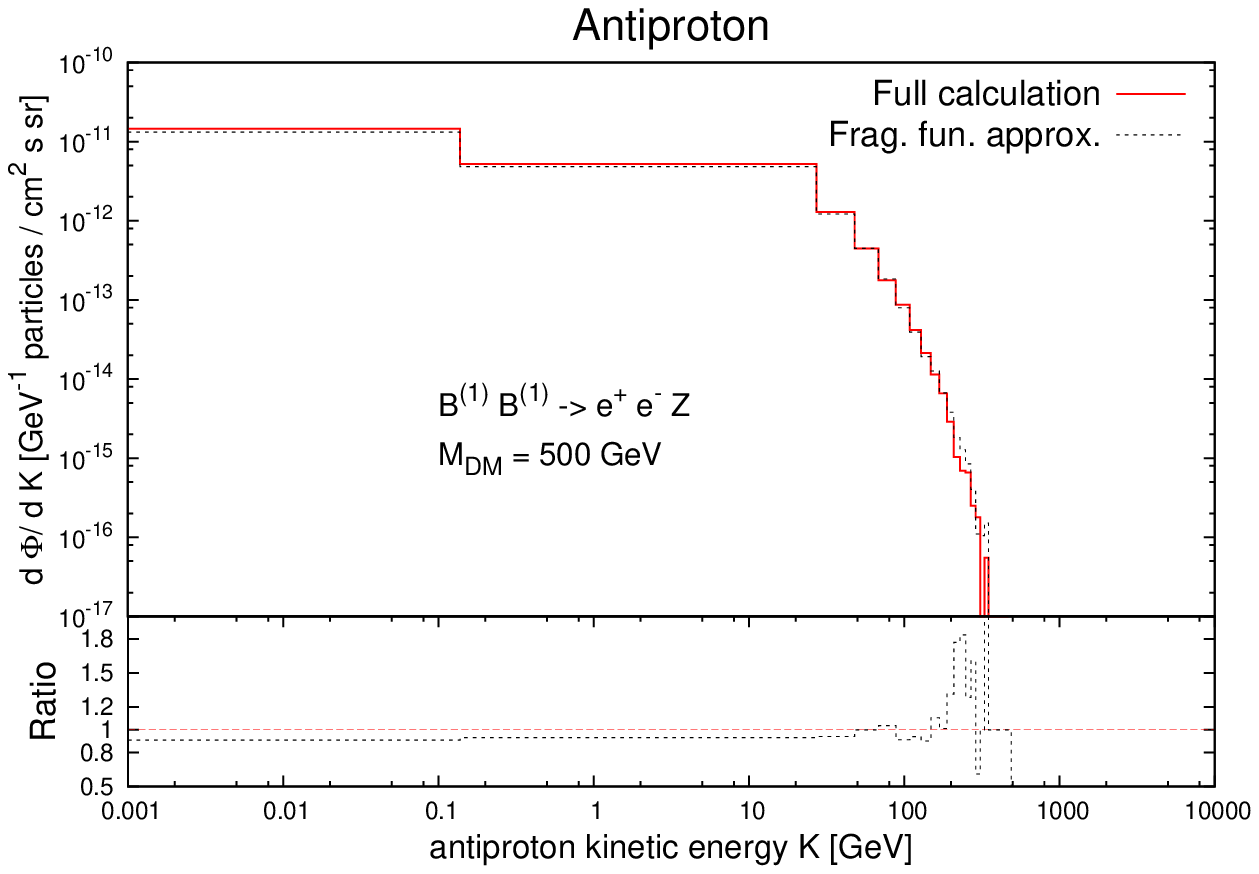}
  \includegraphics[width=0.49\textwidth]{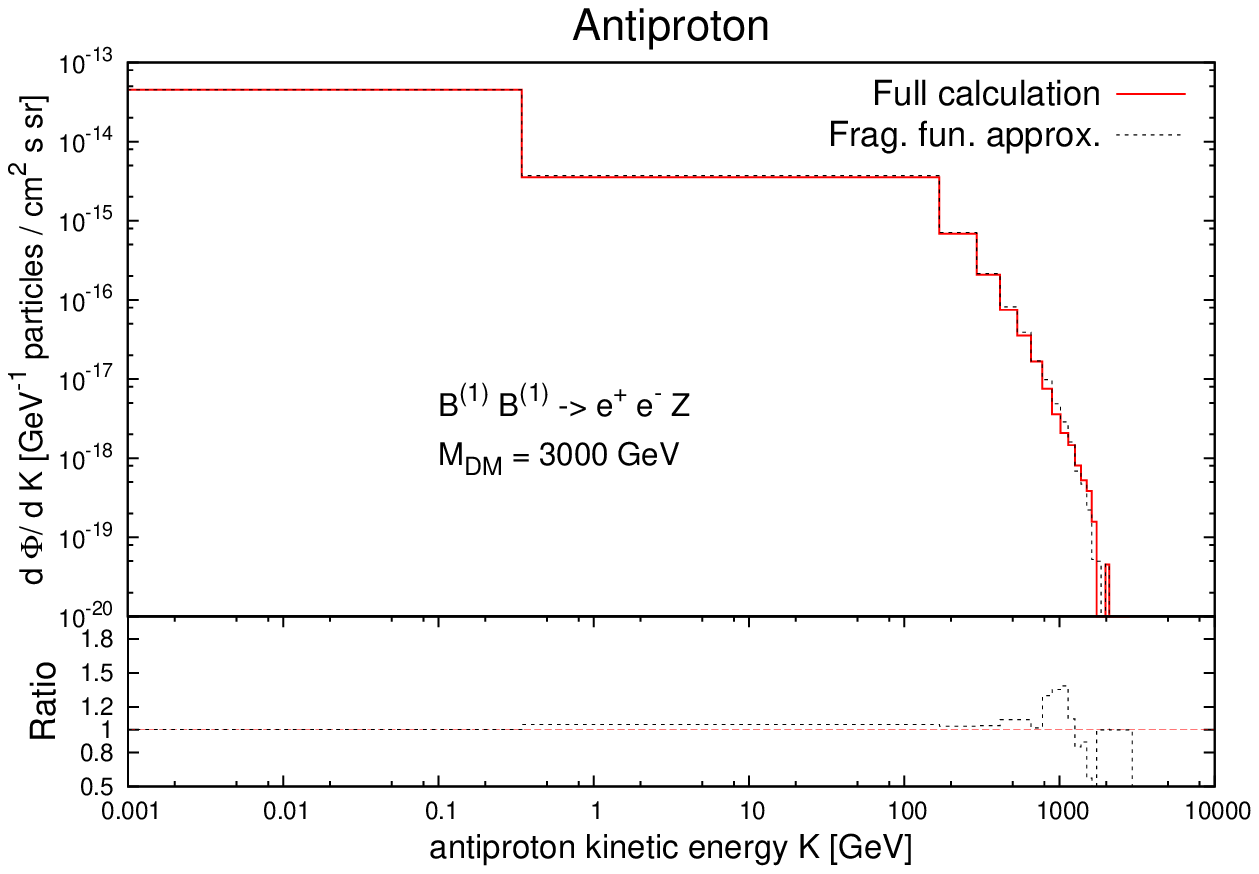}
\caption{Antiproton flux $d\Phi/dK$, Eq.~(\ref{eq:dphidk}), after propagation through the galactic halo for the annihilation of vector dark matter into an electron-positron pair plus a $Z$ boson in a universal extra dimension model. (See the main text for the definition of the model.) Shown is the exact calculation of the flux from $Z$ radiation (red, solid line) and the \ff approximation (black, dotted line) for  $M_{\rm DM} = 500$ and 3000 GeV.}
\label{fig9}
\end{figure}

In our simple leptophilic model set-up, antiprotons are generated exclusively from $Z$-boson decay. As the \ff  provides a good approximation
to the $Z$-boson spectrum of the exact calculation, the flux of antiprotons is also expected to be reproduced well. This is indeed born out by the explicit calculation presented in
Fig.~\ref{fig9}. We find that the exact $(2\to 3)$ calculation and the \ff approach agree within 10\% for $M_{\rm DM} \gtrsim 500$\,GeV.

\section{Conclusions}
\label{ch:conclusions}
The radiation of electroweak gauge bosons may be of crucial importance for dark matter annihilation. Vector boson emission
lifts the helicity suppression of cross sections for Majorana fermion dark matter annihilating into light fermions, and it
alters the spectrum and composition of the secondary flux. The decay of $Z$ bosons, in particular, leads to a secondary flux which
includes all stable SM particles, independent of the model-specific composition of the primary annihilation products.

For heavy dark matter, $M_{\rm DM} \gg m_Z$, the emission of electroweak gauge bosons is enhanced by Sudakov logarithms $\ln^2 (M^2_{\rm DM}/m^2_Z)$. The leading logarithmic contributions from soft and collinear gauge boson emission can be described by electroweak fragmentation functions. We have quantified the quality of the \ff approximation by comparing the \ff prediction with the exact calculation for dark matter annihilation into an electron-positron pair plus a $Z$ boson, ${\rm DM}+{\rm DM} \to e^+e^-+Z$. Specifically, we have investigated
a supersymmetric model and a universal extra dimension model with Majorana fermion and vector dark matter, respectively. We provide predictions for the energy distribution of the $Z$ boson and the secondary flux of positrons and antiprotons after propagation through the galactic halo in both the \ff approximation and the exact calculation of the $(2\to 3)$ process.

We find that the \ff approach fails for the supersymmetric model with Majorana fermion annihilation into electron-positron pairs. The emission of soft/collinear vector bosons, as included in the \ff approximation, is not sufficient to lift the helicity suppression of the lowest-order annihilation cross section. For vector dark matter annihilation, on the other hand, the \ff approach works very well and can describe the shape and normalisation of the secondary flux accurately. Specifically, we find that the particle fluxes obtained from the exact $(2\to 3)$ calculation and the \ff approach agree to better than 10\% for $M_{\rm DM} \approx 500$\,GeV and to better than 2\% for $M_{\rm DM} \approx 1$\,TeV.

Thus, the \ff approximation provides a
simple and model-independent way to obtain realistic predictions for particle fluxes from \DM annihilation for those models where the
annihilation cross section is not suppressed at the lowest order.

\section*{Acknowledgement}

We would like to thank Alexander M\"uck and Torbj\"orn Sj\"ostrand for useful discussions, Ang\'{e}lina Bieler for her collaboration at an early stage of this work,
and Benedikt Marquardt for reading the manuscript. MK is grateful to SLAC and
Stanford University for their hospitality. This work was supported by the Deutsche Forschungsgemeinschaft through the
graduate school ``Particle and Astroparticle Physics in the Light of the LHC'' and through the collaborative research centre TTR9
``Computational Particle Physics'',  and by the U.S.\ Department of Energy under contract DE-AC02-76SF00515. \newpage

\bibliographystyle{unsrt}
\bibliography{EW_dark_matter}

\end{document}